# Feshbach Shape Resonance for High $T_c$ Pairing in Superlattices of Quantum Stripes and Quantum Wells

Antonio Bianconi

*Dipartimento di Fisica, Università di Roma "La Sapienza", P. Aldo Moro 2, 00185 Roma, Italy*

**Abstract**:  The Feshbach shape resonances in the interband pairing in superconducting superlattices of quantum wells or quantum stripes is shown to provide the mechanism for high Tc superconductivity. This mechanism provides the $T_c$ amplification driven by the architecture of material: superlattices of quantum wells (intercalated graphite or diborides) and superlattices of quantum stripes (doped high $T_c$ cuprate perovskites) where the chemical potential is tuned to a Van Hove-Lifshitz singularity (vHs) in the electronic energy spectrum of the superlattice associated with the change of the Fermi surface dimensionality in one of the subbands.

## 1.    INTRODUCTION

The light element diborides $AB_2$ with ($A^{3+}$=Al, $A^{2+}$=Mg) are artificial intermetallics synthesized in the 20th century: first, $AlB_2$ in 1935 [1] followed by $MgB_2$ in 1953 [2,3] as a product of the search of new boron compounds for nuclear reactor bars in the fifties. $MgB_2$ has been available in kilogram-size bottles for 40 years from suppliers of inorganic chemicals as a reagent used in metathesis reactions (in which compounds change partners) and in some commercial preparations of elemental boron [4]. The synthesis of $Al_{1-y}Mg_yB_2$ ternary compounds has been first reported in 1971 [5]. The electronic structure of $Al_{1-y}Mg_yB_2$ is similar to that of hole doped intercalated graphite [6,7]. In $AlB_2$ the Fermi level crosses only the π subband since it is in the partial gap between a filled and an empty σ subband like in graphite. In the ternary intermetallics $Al_{1-y}Mg_yB_2$ by changing the magnesium content y it is possible to dope by holes the band structure of the superlattice of boron layers [8]. The π electrons provide the metallic bonding between the boron layers in the c axis direction. The σ electrons provide the covalent bonding within boron atoms in the a,b planes forming in the graphene-like monolayer a quasi 2D electronic band as in graphite. The σ band is not hybridized with the π electrons or the sp orbital of intercalated ions. The dispersion of the σ sub-band in the c-axis direction is only determined by the electron hopping $t_\perp$ between the graphene-like boron mono-layers that is controlled by their distance (i.e., the c-axis of the $AlB_2$ structure).

The light element diborides, alloys made of low $T_c$, or non superconducting elements, like aluminum and magnesium, have not been suspected to show either low or high $T_c$ superconductivity according with the B.T. Mathias comprehensive review published in 1963 [9] and still valuable today that discussed superconductivity in all the known elements, alloys, and compounds and related the occurrence and non-occurrence of superconductivity to crystal structure as well as to the Matthias rules [10,11].

The electronic structure of transition metal diborides $AB_2$ with A=Nb,Ta,Mn,Cr… is different since the boron σ orbital is hybridized the d orbital of the transition metal ions forming 3D bonding wavefunctions extending in the interlayer space between the boron layers. At low temperature they show superconducting (A=Nb,Ta) [12] or magnetic order (A=Mn,Cr). During the sixties and seventies it was suspected that transition element alloys A=Mn,Cr would be superconducting [13,14], Therefore these intermetallic boride superconductors attracted a large interest being at the borderline between superconductivity and ferromagnetism [9] where high $T_c$ alloys of transition metals and actinides [15] were found. The research was extended to binary hexa- and dode-caborides as the ternary rare-earth rhodium borides where the very interesting reentrant superconductivity and magnetic ordering phenomena were found [16].

$AlB_2$ and $MgB_2$ and other light element diborides were not expected to be superconductors for the standard BCS or other exotic pairing theories involving magnetic interactions, therefore their superconducting properties have not been measured for 48 years. In a patent of 1993 [17,18] and in papers [19,20] we have described a new process to increase the superconducting critical temperature in elements with low $T_c$ and free-electron-like electronic bands. It was shown that the superconducting critical temperature can be amplified by tuning the chemical potential to a Feshbach shape resonance in the interband pairing. This is obtained by controlling the material architecture (a superlattice of metallic layers, as in doped graphite-like layered materials, a superlattice of quantum stripes or wires, as in doped crystals of nanotubes, and a superlattice of quantum dots, as in doped fullerenes). The Fermi level is tuned by changing the charge density. The interband exchange-like pairing has been proposed since 1959 [21-30] as a non BCS pairing that plays a key role in thin films [31], cuprates [32-40], doped fullerenes [41], borocarbides [42], ruthenates [43] and intercalated graphite [44-45] however only rare superconductors are in the requested clean limit.

## 2.    INTERBAND PAIRING

The "shape resonances" (first described by Feshbach in nuclear elastic scattering cross-section for the processes of neutron capture and nuclear fission) deal with the configuration interaction between different excitation channels including quantum superposition of states corresponding to *different spatial locations* [46]. Therefore these resonances could appear under special conditions in the inter-band pairing in superconductors where a pair is transferred between two Fermi surfaces belonging to bands that are not hybridized and are in different spatial locations. Therefore they appear only in multiband superconductors in the clean limit where the disparity and negligible overlap between electron wave-functions of the different bands should suppress the impurity scattering rate for single electrons.

These Feshbach shape resonance (FSR) resonances in the interband pairing are similar to the Feshbach resonances in ultra-cold fermionic gases that are used to raise the ratio $T_c/T_F$ of the superfluid critical temperature $T_c$ on the Fermi temperature $T_F$ [47].

The BCS wave-function of the superconducting ground state has been constructed by configuration interaction of all electron pairs (+k with spin up, and -k with spin down) on the Fermi surface in an energy window that is the energy cut off of the interaction,

$$|\Psi_{BCS}\rangle = \prod_k (u_k + v_k c^+_{k\uparrow} c^+_{-k\downarrow})|0\rangle \qquad (1)$$

where $|0\rangle$ is the vacuum state, and $c^+_{k\uparrow}$ is the creation operator for an electron with momentum k and spin up. The Schrieffer idea [24] of this state with off-diagonal long range order came from the configuration interaction theory by Tomonaga involving a pion condensate around the nucleus.

In anisotropic superconductivity one has to consider configuration interaction between pairs, in an energy window ΔE around the Fermi level, in different locations of the k-space with a different pairing strength, that gives a k-space dependent superfluid order parameter i.e., a k-dependent superconducting gap. A particular case of anisotropic superconductivity is multiband superconductivity, where the order parameter and the excitation gap are mainly different in different bands.

The advances in this field are related with the development of the theory of configuration interaction between different excitation channels in nuclear physics including quantum superposition of states corresponding to *different spatial locations* for interpretation of resonances in nuclear scattering cross-section related with the Fano configuration interaction theory for autoionization processes in atomic physics.

The theory of two band superconductivity, including the configuration interaction of pairs of opposite spin and momentum in the *a*-band and *b*-band, was developed on basis of the Bogolyubov transformations where the many body wave function is given by

$$|\Psi_{Kondo}\rangle = \prod_k (u_k + v_k a^+_{k\uparrow} a^+_{-k\downarrow}) \prod_{k'} (x_{k'} + y_k b^+_{k'\uparrow} b^+_{-k'\downarrow})|0\rangle \qquad (2)$$

The element corresponding to the transfer of a pair from the *a*-band to the *b*-band or vice versa appears with the negative sign [25] in the expression of the energy [23]. This gain of energy is the origin of the increase of the transition temperature driven by interband pairing. The two band superconductivity has been proposed for metallic elements and alloys, for doped cuprate perovskites, for magnesium diboride and for few other materials as Nb doped $SrTiO_3$, $Sr_2RuO_4$, $YNi_2B_2C$, $LuNi_2B_2C$ and $NbSe_2$.

The multiband superconductivity shows up only in the "clean limit", where the single electron mean free path for the interband impurity scattering satisfies the condition $l > \hbar v_F / \Delta_{av}$ where $v_F$ is the Fermi velocity and $\Delta_{av}$ is the average superconducting gap. Therefore the criterium that the mean free path should be larger than the superconducting coherence length must be met. This is a very strict condition that implies also that the impurity interband scattering rate $\gamma_{ab}$ should be very small $\gamma_{ab} << (1/2)(K_B/\hbar)T_c$.

Therefore most of the metals are in the "dirty limit" where the interband impurity scattering mixes the electron wave functions of electrons on different spots on bare Fermi surfaces and it reduces the system to an effective single Fermi surface.

The "interchannel pairing" or "interband pairing" that transfers a pair from the "a"-band to the "b"-band and viceversa in the multiband superconductivity theory is expressed by

$$\sum_{k,k'} J(k,k')(a^+_{k\uparrow} a^+_{-k\downarrow} b_{-k\downarrow} b_{k\uparrow}) \qquad (3)$$

where $a^+$ and $b^+$ are creation operators of electrons in the "a" and "b" band respectively and $J(k,k')$ is an exchange-like integral. This interband pairing interaction may be repulsive as it was first noticed by Kondo. Therefore it is a non-BCS pairing process since in the BCS theory an attractive interaction is required for the formation of Cooper pairs. Another characteristic feature of multiband superconductivity is that the order parameter shows the sign reversal in the case of a repulsive interband pairing interaction.

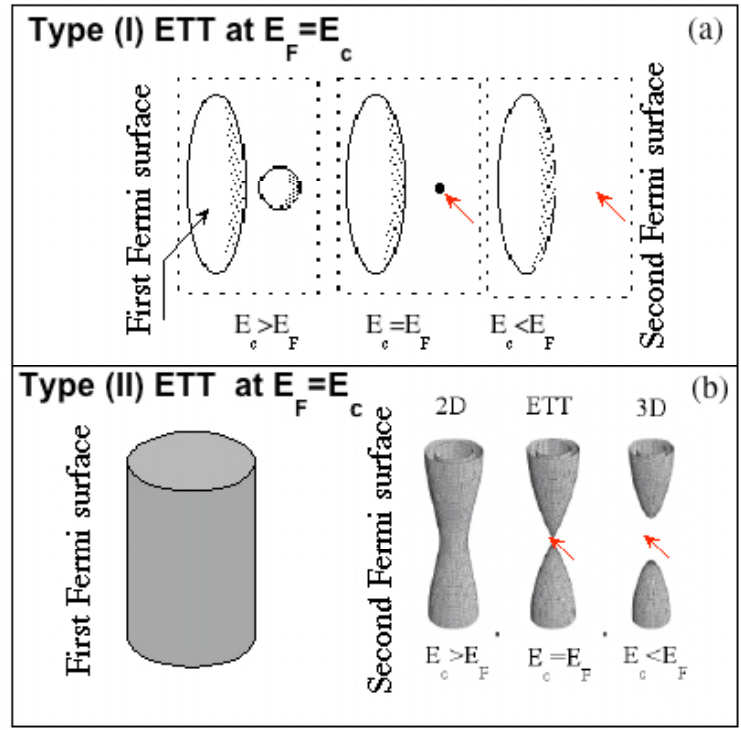

*Fig:1*. The different types of Lifshitz electronic topological transition (ETT): The upper panel shows the type (I) ETT where the chemical potential $E_F$ is tuned to a Van Hove singularity (vHs) at the bottom (or at the top) of a second band with the appearance (or disappearance) of a new detached Fermi surface region. The lower panel shows the type (II) ETT with the disruption (or formation) of a "neck" in a second Fermi surface where the chemical potential $E_F$ is tuned at a vHs associated with the gradual transformation of the second Fermi surface from a two-dimensional (2D) cylinder to a closed surface with three dimensional (3D) topology characteristics of a superlattice of quantum wells.

The non-BCS nature of the interband pairing process is indicated also by the fact that, when it is dominant, the isotope effect vanishes even if the intra-band attractive interaction in each band is due to the electron-phonon coupling. Moreover the effective repulsive Coulomb pseudopotential in the Migdal-Eliashberg theory goes to zero (so the effective coupling strength increases) where the interband pairing is dominant. In this work we will discuss the particular case of multiband superconductivity where a Van Hove-Lifshitz feature in the electronic energy spectrum shown in Fig. 1 occurs within the energy window for the pairing in one of the subbands. We will describe that a resonance in the exchange-like interband pairing (called shape resonance of Feshbach resonance) occurs where the chemical potential $E_F$ is tuned near a type I or type II electronic topological transitions.

## 3. SHAPE RESONANCES OR FESHBACH RESONANCES IN INTERBAND PAIRING

### 3.1 The case of type I ETT

The process for increasing $T_c$ by a Feshbach "shape resonance" was first proposed by Blatt and Thompson [31] in 1963 for a superconducting thin film. Blatt considered the ideal case of a very thin membrane (see Fig. 2) made of a superconducting material where the thickness L of the membrane is so small that it is of the order of the wavelength $\lambda_F$ of the electrons at the Fermi level.

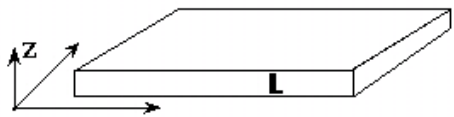

*Fig: 2.* The pictorial view of a single superconducting film of thickness L.

In this ideal film the wavevector of the electrons in the direction normal to the film surface is quantized in units of $\pi/L$ as it is shown in Fig. 3. Therefore we get two different subbands: the first one is characterized by $k_{z1}=\pi/L$ and the second one by $k_{z2}=2\pi/L$ as shown in Fig. 4.
It is possible to change the energy of the energy $E_2$ of the bottom of the second subband by changing the film thickness L and therefore the energy difference $z=E_F-E_2$ that is the physical parameter controlling the distance from the quantum critical point where the Fermi surface of the system shows a topological change, known as an Electronic Topological Transition (ETT) of type I shown in Fig. 5. The thickness L is tuned in such a way that it is so small that $k_F$ is of the order of $k_{z2}=2\pi/L$. Therefore the Fermi level $E_F$ can be driven close to the bottom of the second subband $E_2$ as shown in Fig. 3.

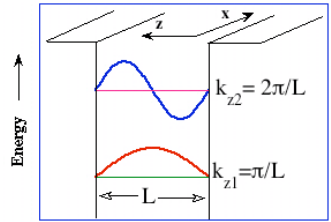

*Fig: 3.* The electrons in a single superconducting film of thickness L are confined in a quantum well of size L. The wave-vector of the electron in the direction normal to the film surface is quantized in units of $\pi/L$. Therefore we get two different subbands characterized by $k_{z1}=\pi/L$ and $k_{z2}=2\pi/L$ respectively.

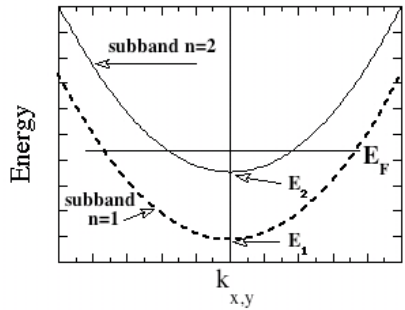

*Fig: 4.* The two different subbands for the electron gas in a single superconducting film of thickness L: the first n=1 is characterized by $k_{z1}=\pi/L$ and the second one, n=2, by $k_{z2}=2\pi/L$. The bottom of the first and second subbands are at $E_1$ and $E_2$ respectively.

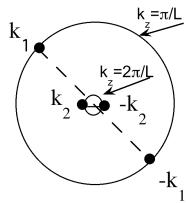

*Fig:5.* The large and small Fermi surfaces for a thin film of thickness L where the Fermi energy $E_F>E_2$. The small Fermi surface disappears for $E_F<E_2$. The Fermi level can be tuned in the energy range around the bottom of the second subband at $E_2$ by changing the film thickness or the charge density. At this quantum critical point the 2D Fermi surface of the second subband appears or disappears and the group velocity of the electrons in second subband is very slow approaching zero for $E_F=E_2$.

Blatt considered only the case of the physical parameter z tuning driven by the change of the film thickness but the ETT parameter z can be controlled also by tuning $E_F$ by the variation of the charge density in the metallic film. The ETT occurs at z=0, in fact going from negative to positive value of the Lifshitz parameter z the second Fermi small surface appears. For $E>E_2$ we have two different closed Fermi surfaces with different parity as shown in Fig. 5.

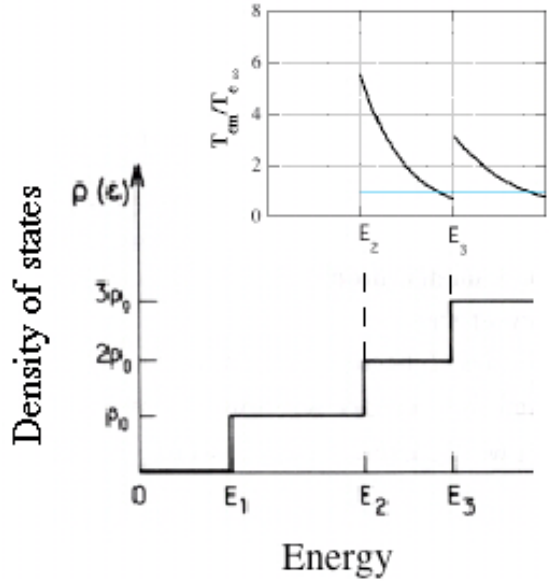

*Fig: 6.* The total density of states (DOS) of a single film of thickness L. The partial DOS due to each subband $\rho_0$ is constant as a function of the energy so the total DOS shows sharp steps at the bottom of the subbands $E_1$, $E_2$ and $E_3$ respectively. The calculation of the superconducting critical temperature including the interband pairing shows the resonance due to the shape resonance or Feshbach resonance near the bottom of the second subband, in the range $E_F \geq E_2$, and a similar resonance in the range $E_F \geq E_3$.

The first circle is a large Fermi surface due to the first subband where electrons have a large kinetic energy and large wave-vector $k_1$; the second circle is a small Fermi surface where the electrons have a small kinetic energy and small wave-vectors $k_2$. By going toward z=0 the group velocity of the electrons in the second subband goes to zero. Therefore approaching the ETT the interband pairing between the two subbands concerns a quantum transfer of pair $(k_1,-k_1)$, made by pairing particles with high kinetic energy in the BCS regime, into a pair $(k_2,-k_2)$ made by pairing particles with low kinetic energy in the Bose regime where z is of the order of the energy cut off of the interaction. Under these condition the superconducting temperature shows an amplification driven by the exchange like pairing that is called a shape resonance of Feshbach resonance as shown in Fig. 6.

It has been shown here that there is a Feshbach resonance in the off diagonal terms of the interaction by tuning the chemical potential near the critical energy $E_F=E_n$ for a Lifshitz electronic topological transition (ETT) of the type (I) shown in Fig. 1, This first type of Feshbach shape resonance occurs by tuning the Lifshitz parameter $z=E_F-E_n$ around z=0. At this ETT a small Fermi surface of a second subband disappears while the large 2D Fermi surface of a first subband shows minor variations.

Superconductivity here is in the "clean limit" since the single electrons cannot be scattered from one to the other band because of different parity of the subbands but configuration interaction between pairs in different bands is possible in an energy window around $E_F=E_n$. In the sixties the proposal of Blatt and Thompsom was rejected by the scientific community since phase fluctuations due to the confinement of the electron gas in two dimension are expected to suppress the superconducting critical temperature.

### 3.2 The case of a type II ETT

In 1993 we have proposed a different scenario for the shape resonance focusing on the Feshbach resonance in proximity of a type II ETT that could be realized in a superlattice of membranes. In this case the basic system is a superlattice of metallic membranes of thickness L intercalated by a different material of thickness W forming a superlattice of the period $\lambda =L+W$ as shown in Fig. 7..

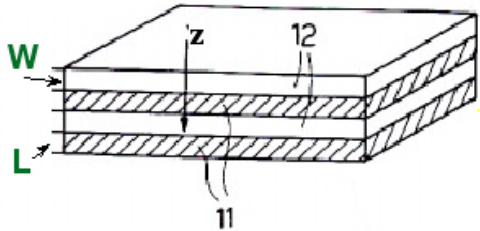

*Fig 7.* A superlattice of metallic membranes of thickness L (dashed layers) intercalated by a different material of thickness W forming a superlattice of the period $\lambda =L+W$ in the z direction.

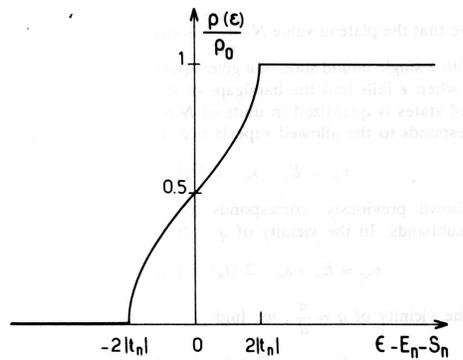

*Fig 8.* The Density of States (DOS) in the energy range near the bottom of a subband as a function of the energy distance from the bottom of a 2D subband. The sharp step at $\varepsilon=0$ for a single layer shows a broadening in an energy window of $D=4t_n$ where $t_n$ is the hopping energy between neighbour layers in the normal direction. At the energy $\varepsilon=+2t_n$ there is a 2D/3D ETT of type II while the bottom of the band is shifted to the energy $\varepsilon =-2t_n$ where there is a ETT of type I.

In this case, first described in 1993, the Fermi level is tuned near the bottom of the second or third subband of the superlattice. The density of states near the bottom of each subband for a superlattice does not shows a sharp step as for the single membrane, but a smooth increase due the presence of novel regime where the Fermi surface has a three dimensional (3D) topology as it is shown in Fig, 9. We assume in Fig. 8 that the sharp step of the DOS at the bottom of a subband of a single quantum well occurs at ε=0. The finite hopping energy $t_n$ between neighbour layers in the normal direction of a superlattice of quantum wells shift the bottom of the subband energy ε =-2$t_n$ there is a ETT of type I. In the energy range D=4$t_n$ the DOS shows a smooth increase up to the energy ε=+2$t_n$ and the Fermi surface has a three dimensional (3D) closed topology. At the critical energy $E_e$=+2$t_n$ The Fermi surface shows a 3D/2D ETT of type II in fact beyond this energy the Fermi surface has a cylindrical shape with 2D dimensionality.

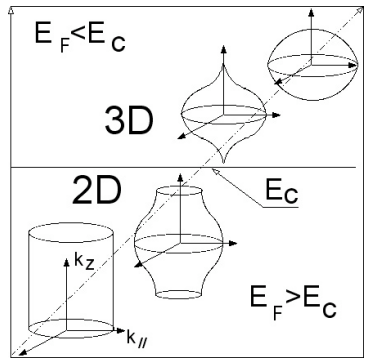

*Fig: 9.* The cylindrical 2D Fermi surface becomes a closed 3D Fermi surface as the chemical potential crosses the energy $E_c$ for the 2D/3D ETT.

The prediction that by tuning the Fermi level at a 2D/3D ETT in superlattice of quantum wells the critical temperature will be amplified by the shape resonance in the interband pairing has been verified experimentally in 2001 for the case of superconductivity in magnesium diboride.

## 4.       FESHBACH SHAPE RESONANCE IN DIBORIDES

The idea that the Feshbach shape resonance in a superlattice of quantum wells increases the critical temperature has been well verified experimentally in these last 5 years [48-58] the diborides are made of a superlattice of atomic boron monolayers intercalated by different atoms as Al or Mg as shown in Fig. 10. The interacalated atoms change the period of the superlattice in the z direction and change the charge density in the boron monlayers due to the charge transfer. In $Al_{1-y}Mg_yB_2$ the position of the Fermi level is tuned toward the filled σ sub-band by Mg for Al substitution. For y=0 the Fermi surface if like that of graphite where only the large π band crosses the Fermi level. For $y \approx 0.44$ the Fermi level reaches the top of the σ subband and for 0.44<y<0.7 the large π Fermi surface typical of graphite coexists with a second small hole-like 3D σ Fermi surface that for 0.7<y<1 has a 2D topology. Therefore the electronic structure of $Al_{1-y}Mg_yB_2$ for y>0.44 is like that of a heavily hole doped graphite where the additional σ subband appears at the Fermi level so that the conduction electron Fermi gas is made of two components: the σ and π electrons having two different spatial locations (in the graphene-layers and in the interlayer space between them respectively) characterized by the disparity between their wavefunction. The disparity and the negligible overlap between electron wave functions in the different σ and π subbands suppress both the hybridization between the two components and the single-electron interband impurity scattering rate that makes possible multiband superconductivity with a key role of the interband pairing. Since the Feshbach shape resonance (FSR) in the interband pairing occurs where there is a change of the Fermi surface topology, i.e., an electronic topological transition (ETT) in one of the subbands it has been proposed that the high $T_c$ in magnesium diboride is determined by the proximity to the FSR in the interband pairing term centered at the 2D to 3D ETT in the σ Fermi surface and the superconducting properties of $Al_{1-y}Mg_yB_2$ can be assigned only to band filling effects [48-53] with minor effects of impurity scattering.

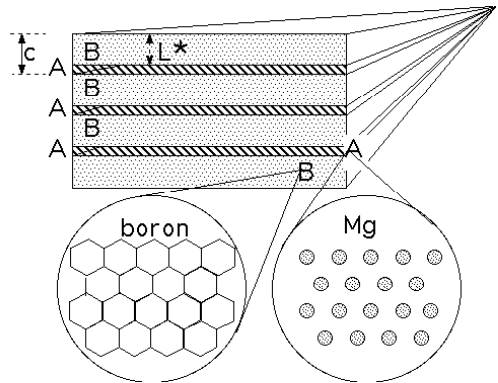

*Fig:10.* For the electron gas in a single superconducting film of thickness L we get two different subbands: the first one characterized by $k_{z1}=\pi/L$ and the second one by $k_{z2}=2\pi/L$. The bottom of the bands are at $E_1$ and $E_2$ respectively.

Recently several experiments have proven that the aluminium [54,55], carbon [55,56] and scandium [57,58] substitution for magnesium tune the chemical potential, with minor effect of the impurity scattering keeping the superconductor in the clean limit. At y=0.7 in $Al_{1-y}Mg_yB_2$ there is a crossover in the hierarchy superconducting gap energies in the σ and π bands [53].

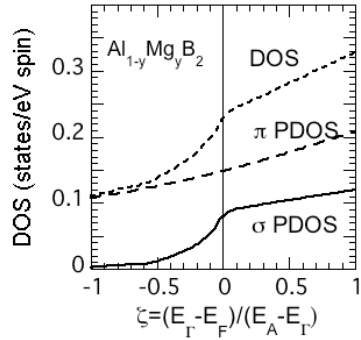

*Fig.:11*. The total density of states (DOS) and the partial DOS (PDOS) of the σ and π band at the Fermi level $E_F$ in $Al_{1-y}Mg_yB_2$ as a function of the reduced Lifshitz parameter ζ where $E_A$ is the energy of the top of the σ band at the A point in the band structure, and $E_\Gamma$ is where the σ Fermi surface changes from a 2D corrugated tube for $E_\Gamma > E_F$ to a closed 3D Fermi surface for $E_\Gamma < E_F$. The type (I) electronic topological transition (ETT) with the appearance of the closed 3D σ Fermi surface occurs by tuning the Fermi energy at the critical point $E_F = E_A$ (ζ=-1). The type (II) ETT with the disruption of a "neck" in the σ Fermi surface occurs by tuning the chemical potential $E_F$ at the critical point in the band structure $E_F = E_\Gamma$ (ζ=0) while the large π Fermi surface keeps its 3D topology.

Therefore for y>0.7 the σ gap is the largest but for 0.4<y<0.7 the system is like a superconducting hole doped graphite with the main gap in the π(3D) band and a smaller gap in the small pocket of the σ(3D) band. In this work we have investigated the variation of the superconducting properties as a function of y in $Al_{1-y}Mg_yB_2$ in order to investigate this interesting superconducting phase.

In $MgB_2$ the σ band is partially unoccupied due to the electron transfer from the boron layers to the magnesium layers. In fact the chemical potential $E_F$ in $MgB_2$ is at about 750 meV below the energy $E_A$ of the top of the σ band. Moreover the chemical potential $E_F$ in $MgB_2$ is also at about 350 meV below the energy of the Γ point in the band structure ($E_\Gamma > E_F$). Therefore the σ Fermi surface of $MgB_2$ where $E_F < E_\Gamma < E_A$ has the corrugated tubular shape with a two-dimensional topology for the case $E_F < E_\Gamma$. Going from below ($E_F < E_\Gamma < E_A$) to above the energy of the Γ point the σ Fermi surface becomes a closed Fermi surface with 3D topology like in $AlMgB_4$, y=0.5, that belongs to the Fermi surface type for the case $E_\Gamma < E_F < E_A$.

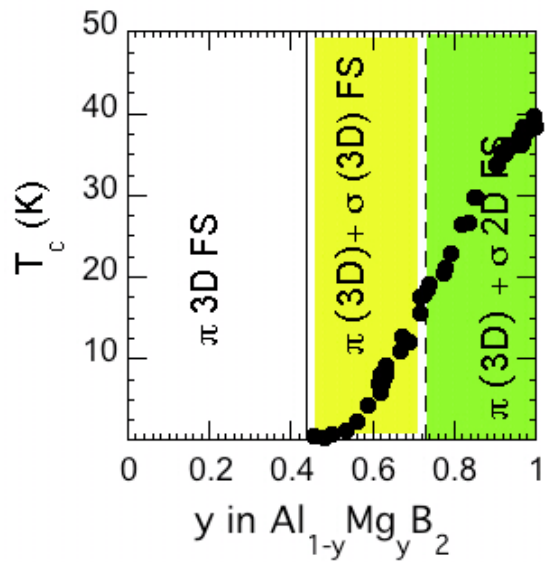

*Figure :12.* The superconducting critical temperature as a function of the magnesium content y.

Therefore by tuning the chemical potential $E_F$ it is possible to reach the point where $E_F$ is tuned at the 2D/3D ETT at y=0.7. This is a type (II) 2.5 Lifshitz electronic topological transition (ETT) with the disruption of a "neck" in the σ Fermi surface with the critical point at $E_F=E_\Gamma$. The changes of physical properties near the ETT transition are studied here as a function of the reduced Lifshitz parameter $\zeta=(E_\Gamma-E_F)/(E_A-E_\Gamma)$, where $D=(E_A-E_\Gamma)$ is the energy dispersion in the c-axis direction due to electron hopping between the boron layers (D=0.4 eV in $MgB_2$ but it changes with chemical substitutions). The influence of the proximity to a type (II) electronic topological transition on the anomalous electronic and lattice properties of $MgB_2$ is shown by the anomalous behaviour of the Raman spectra.

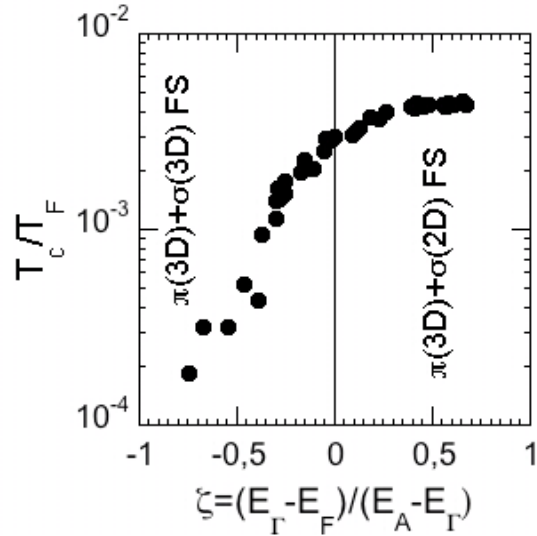

*Fig:13.* The ratio $T_c/T_F$ of the critical temperature $T_c$ and the Fermi temperature $T_F=E_F/K_B$ for the holes in the σ subband in magnesium for aluminum substituted diborides as a function of the reduced Lifshitz parameter ζ in $Al_{1-y}Mg_yB_2$.

Fig. 13 shows the response of the superconducting critical temperature as a function of y in magnesium for aluminum substituted diborides $Al_{1-y}Mg_yB_2$. Fig. 13 shows the ratio $T_c/T_F$ of the critical temperature $T_c$ and the Fermi temperature $T_F=E_F/K_B$ for the holes in the σ subband as a function of the reduced Lifshitz parameter ζ.
.

In conclusion we have reported evidence for the Feshbach shape resonance around a ETT in the particular case of doped $MgB_2$, a multiband superconductor in the clean limit. [58-61].

In high Tc cuprate perovskites it has been shown recently [62-63] that the angular resolved photoemission and local probes show evidence for Fermi surface arcs driven by a superlattice of quantum wires that provide the condition for the Feshbach shape resonance.

There is now a relevant effort toward manipulation of the nanoscale texture by controlling frustrated first order phase transition in order to get modulated heterostructure at atomic limit that could be the next generation of high temperature superconductors working at room temperature.

## ACKNOWLEDGMENTS

This work is supported by European STREP project 517039 "Controlling Mesoscopic Phase Separation" (COMEPHS).

## REFERENCES

1. W. Hoffmann, and W. Jänicke, *Naturwiss* 23, 851 (1935).
2. V. Russel, R. Hirst, F. Kanda and A. King, *Acta Cryst.* 6, 870 (1953).
3. E. Jones and B. Marsh *J Am. Chem. Soc.* 76, 1434 (1954).
4. E. G. Killian, and R. B. Kaner, Chem. Mater. 8, 333 (1996).
5. N.V. Vekshina, L. Ya. Markovskii, Yu. D. Kondrashev, and T.K. Voevodskyaya, *Zh. Prikl. Khim.* (Leningrad) 44 958-963 (1971)
6. M.S. Dresselhaus, G. Dresselhaus, *Adv. in Phys.* 30, 139 (1981).
7. M.S. Dresselhaus, G. Dresselhaus and P.C. Eklund, *Science of Fullerenes and Carbon Nanotubes,* (Academic Press Inc. San Diego 1996).
8. O. de la Pena, A. Aguayo, and R. de Coss, *Phys. Rev. B* 66, 012511 (2002).
9. B. T. Matthias T. H. Geballe and V. B. Compton *Rev. Mod. Phys.* 35:1. Errata *Rev. Mod. Phys.* 35, 414 (1963).
10. B. T. Matthias *Phys. Rev.* 97, 74 (1955).
11. B. T. Matthias *Superconductivity in the periodic system* in *Progress in Low Temperature Physics*, ed. C. J. Gorter, p. 138. Amsterdam: North-Holland Publishing Co. (1957).
12. L. Leyarovska *et al. J. Less-common Metals* 67, 249 (1979).
13. L. Andersson *et al*. *Solid State Communications* 4, 77 (1966).
14. J. Castaing *et al. J. Phys.Chem.Solids* 33, 533 (1972).
15. J.L. Smith and E.A. Kmetko *J. Less Common Metals* 90, 83 (1983); J. L. Smith and R.S. Riseborough *Journal of Magnetism and Magnetic Materials* 47&48 , 545 (1985).
16. B. T. Matthias D. C. Johnston, W. A. Fertig, and M. B. Maple. *Sol. St. Comm.* 26, 141 (1978).
17. A. Bianconi *"Process of increasing the critical temperature $T_c$ of a bulk superconductor by making metal heterostructures at the atomic limit"* United State Patent: No.:US 6, 265, 019 B1. (priority date: 07 12 1993).
18. A. Bianconi "High $T_c$ superconductors made by metal heterostuctures at the atomic limit" European Patent N. 0733271 (priority date: 07 12 1993)
19. A. Bianconi, *Sol. State Commun.* 89. 933 (1994)
20. A. Bianconi, A. Valletta, A. Perali, N. L. Saini *Physica C* 296, 269 - 280 (1998).
21. H. Suhl, B.T. Matthias, and L.R. Walker *Phys. Rev. Lett.* **3**, 552 (1959).
22. V.A. Moskalenko, *Phys. Met. and Metallog.* 8, 25 (1959).
23. J. Kondo *Prog. Theor. Phys.* 29, 1 (1963).
24. J.R. Schrieffer, *Theory of Superconductivity*, 1964, (for interband pairing see 300 pp.).
25. A. J. Leggett, *Prog. Theor. Phys.* 36, 901 (1966); 36, 931 (1966).
26. B.T. Geilikman, R. O. Zaitsev, and V. Z. Kresin, *Sov. Phys. Solid State* 9, 642 (1967).
27. V.A. Moskalenko and M. E. Palistrant, *Sov. Phys. JETP* 22, 536 (1966).
28. L.Z. Kon, *Phys. Met. Metallogr.* (USSR). 23, (1967).
29. P. Entel, and M. Peter, *Jour. of Low Temp. Phys.* 22, 613 (1976).
30. N. Schopohl and K. Scharnberg *Solid State Commun.* 22, 371 (1977).
31. J.M. Blatt and C.J. Thompson *Phys. Rev. Lett.* 10, 332 (1963); C.J. Thompson and J.M. Blatt, *Phys. Lett*. 5, 6 (1963); J. M. Blatt in *Theory of Superconductivity* Academic Press New York 1964, (pag. 362 and 215).
32. D.H. Lee, and J. Ihm, *Solid State Commun.* 62, 811 (1987).
33. K. Yamaij, and A. Abe, *J. Phys. Soc. Japan* 56, 4237 (1987).
34. V.Z. Kresin, and S. A. Wolf, *Phys. Rev. B* 46, 6458 (1992); *Phys. Rev. B* 51, 1229 (1995).
35. N. Bulut, D. J. Scalapino, and R.T. Scalettar, *Phys. Rev. B* 45, 5577 (1992).
36. B.K. Chakraverty, *Phys Rev B* 48, 4047 (1993).
37. R. Combescot, and X. Leyronas, *Phys. Rev. Lett.* 75, 3732 (1995).
38. N. Kristoffel, P. Konsin, and T. Ord, *Rivista Nuovo Cimento* 17, 1 (1994).
39. J.F. Annett, and S. Kruchinin *"New Trends in Superconductivity"* Kluwer Academic Publishers, Dodrecht, Netherlands (2002).
40. A. Bianconi *J. Phys. Chem. Sol.* 67, 566 (2006).
41. R. Friedberg, and T.D. Lee, *Phys. Rev.* B40, 6745 (1989).

42. S.V. Shulga, S.L. Drechsler, G. Fuchs, K.H. Muller, K. Winzer, M. Heinecke and K. Krug, *Phys. Rev. Lett.* 80, 1730 (1998).
43. M.E. Zhitomirsky, and T. M. Rice, *Phys. Rev. Lett.* 87, 057001 (2001).
44. G. Cs´anyi, P. B. Littlewood, A. H. Nevidomskyy, C. J. Pickard and B. D. Simons arXiv:cond-mat/0503569 (23 Mar 2005).
45. T. E. Weller, M. Ellerby, S. S. Saxena, R. P. Smith, N. T. Skipper cond-mat/0503570 (23 Mar 2005).
46. A. Bianconi, and M. Filippi in *" Symmetry and Heterogeneity in high temperature superconductors"* A. Bianconi (ed.) Spinger, Dordrecht, The Netherlands. Nato Sciente Series II Mathematics, Physics and Chemistry vol. 214, pp. 21-53, (2006).
47. C. A. Regal, M. Greiner, and D. S. Jin, *Phys. Rev. Lett.* 92, 040403 (2004).
48. A. Bianconi et al., *cond-mat/0102410* (22 Feb 2001); A. Bianconi et al., *cond-mat/0103211* (9 March 2001)
49. A. Bianconi, D. Di Castro, S. Agrestini, G. Campi, N.L. Saini, A. Saccone, S. De Negri, M. Giovannini, Presented at *APS March meeting,* March 12, 2001 Seattle, Washington, *"Session on $MgB_2$"* tak 79, http://www.aps.org/MAR01/mgb2/talks.html#talks79
50. A. Bianconi, D. Di Castro, S. Agrestini, G. Campi, N. L. Saini, A. Saccone. S. De Negri, M. Giovannini, *J. Phys.: Condens. Matter* 13, 7383 (2001).
51. A. Bianconi, S. Agrestini, D. Di Castro, G. Campi, G. Zangari, N.L. Saini, A. Saccone, S. De Negri, M. Giovannini, G. Profeta, A. Continenza, G. Satta, S. Massidda, A. Cassetta, A. Pifferi and M. Colapietro, *Phys. Rev. B* 65, 174515 (2002).
52. A. Bussmann-Holder and A. Bianconi, *Phys. Rev.* B 67, 132509 (2003).
53. G.A. Ummarino, R.S.Gonnelli, S.Massidda and A.Bianconi, *Physica C* 407, 121 (2004).
54. L. D. Cooley A. J. Zambano, A.R. Moodenbaugh, R. F. Klie, Jin-Cheng Zheng, and Yimei Zhu *Phys. Rev. Lett.* 95, 267002 (2005)
55. P. Samuely, P. Szabó, P.C. Canfield, S.L. Bud'ko *Phys, Rev. Lett.* 95, 099702 (2005).
56. S. Tsuda, T. Yokoya, T. Kiss, T. Shimojima, S. Shin, T. Togashi, S. Watanabe, C. Zhang, C. T. Chen, S. Lee, H. Uchiyama, S. Tajima, N. Nakai, and K. Machida *Phys. Rev. B* 72, 064527 (2005).S. Tsuda et al. cond-mat/0409219.
57. S. Agrestini, C. Metallo, M. Filippi, L. Simonelli, G. Campi, C. Sanipoli, E. Liarokapis, S. De Negri, M. Giovannini, A. Saccone, A. Latini, and A. Bianconi, *Phys. Rev. B* 70, 134514 (2004).
58. A. Bianconi *Journal of Superconductivity* **18**, 25 (2005).
59. A. Bianconi, *Symmetry and Heterogeneity in high temperature superconductors"* Spinger The Netherlands (2006). ISBN-10 1-4020-3987-5
60. M. Filippi, A. Bianconi and A. Bussmann-Holder *J. Phys. IV* France 131, 49 (2005).
61. G. G. N. Angilella, A. Bianconi, and R. Pucci *Journal of Superconductivity* 18, 19 (2005)
62. A. Bianconi, S. Agrestini, G. Campi, M. Filippi, and N. L. Saini, *Current Applied Phys.* 5, 254 (2005)
63. Antonio Bianconi *J. of Physics and Chemistry of Solids* 67, 566 (2006).